\documentstyle[prl,aps,floats,twocolumn,epsfig]{revtex}

\newcommand{\beq}{\begin{equation}}
\newcommand{\eeq}{\end{equation}}
\newcommand{\bea}{\begin{eqnarray}}
\newcommand{\eea}{\end{eqnarray}}

\def\laq{\raise 0.4ex\hbox{$<$}\kern -0.8em\lower 0.62
ex\hbox{$\sim$}}
\def\gaq{\raise 0.4ex\hbox{$>$}\kern -0.7em\lower 0.62
ex\hbox{$\sim$}}

\def \ra {\rightarrow}

\def \b {\beta}

\begin{document}

\par
\begingroup
\twocolumn[%

\begin{flushright}
BA-TH/99-332\\
gr-qc/9902060\\
To appear in {\bf Phys. Rev. D}
\end{flushright}
\bigskip

{\large\bf\centering\ignorespaces
Inflation and Initial Conditions in the Pre-Big Bang Scenario
\vskip2.5pt}
{\dimen0=-\prevdepth \advance\dimen0 by23pt
\nointerlineskip \rm\centering
\vrule height\dimen0 width0pt\relax\ignorespaces
M. Gasperini 
\par}
{\small\it\centering\ignorespaces
Dipartimento di Fisica, Universit\`a di Bari, 
Via G. Amendola 173, 70126 Bari, Italy\\
and Istituto Nazionale di Fisica Nucleare, Sezione di Bari, Bari,
Italy 
\par}

\par
\bgroup
\leftskip=0.10753\textwidth \rightskip\leftskip
\dimen0=-\prevdepth \advance\dimen0 by17.5pt 
\nointerlineskip
\small\vrule width 0pt height\dimen0 \relax

The pre-big bang scenario describes the evolution of the
Universe from an initial state approaching the flat, cold, empty, 
string perturbative vacuum. The choice of such an initial state is
suggested by the present state of our Universe if we accept that the
cosmological evolution is (at least partially) duality-symmetric.
Recently, the  initial conditions of the pre-big bang scenario have
been criticized  as they introduce large dimensionless parameters
allowing the Universe to be ``exponentially large from the very
beginning". We agree that a set of initial parameters (such
as the initial homogeneity scale, the initial entropy) larger than 
those determined by the initial horizon scale, $H^{-1}$,   would be
somewhat unnatural to start with. However, in the pre-big bang
scenario, the initial parameters are all bounded by the size of the
initial horizon.  The basic question thus becomes: is a maximal
homogeneity scale of order $H^{-1}$ necessarily unnatural  if the
initial curvature is small and, consequently, $H^{-1}$ is very  large in
Planck (or string) units?  In the impossibility of experimental
information one could exclude ``a priori", for  large horizons,  the
maximal homogeneity scale $H^{-1}$ as a natural initial condition. In
the pre-big bang scenario, however, pre-Planckian initial conditions are
not necessarily washed out by inflation and are accessible (in principle)
to observational tests, so that their  naturalness could be also analyzed
with a Bayesan approach, in terms of ``a posteriori" probabilities. 

\par\egroup
\vskip2pc]
\thispagestyle{plain}
\endgroup

Recently, the validity of the pre-big bang scenario as a viable
inflationary model has been questioned on the grounds of its
initial conditions \cite{1}. 

The main criticism raised against models in which the Universe 
evolves from the flat, zero-interactions, string perturbative vacuum
is mainly based on two points \cite{1}. The first concerns the
homogeneity problem, in particular the largeness of the initial
homogeneous region in string (or Planckian)  units; the second
concerns the flatness problem, and in particular the two large
dimensionless parameters  (the inverse of the string coupling and of
the curvature, in string units) characterizing the Universe at the
beginning of inflation. The fact that, as a consequence of these large
numbers, ``the pre-big bang Universe must be very huge and
homogeneous from the very beginning" is quoted as a serious
problem, supporting the conclusion that ``the current version of the
pre-big bang scenario cannot replace usual inflation" \cite{1}.

I agree with the remarks concerning the initial size of the 
Universe (indeed, the need for an initial state with
a Universe very large in Planck units was already noted in the
first paper on the pre-big bang scenario \cite{1a} and, even before, in
the context of string-driven superinflation \cite{2}; in
particular, the condition  on the duration of inflation, reported in
\cite{1} as eq. (8),  was already derived in \cite{3}).  
The large initial size of the Universe is only part of the conditions to 
be imposed at the onset of pre-big bang inflation, and I also agree with
the fact that a successful pre-big bang  scenario requires
an initial state characterized by very small (or very large)
dimensionless ratios measuring the initial curvature and coupling
constant, and possibly leading to a fine-tuning problem, as first pointed
out in \cite{4}. I disagree, however, with the conclusion presented in
\cite{1},  and I would like to point out some arguments, hoping to
clarify a different point of view on a large initial Universe. 

I will concentrate, in particular, on the largeness of the initial
horizon scale, which can be thought to be  at the ground of the
various objections discussed in \cite{1}. The large
dimensionless ratios of the initial state, when referred to the
Einstein frame in which the Planck length is fixed, 
correspond indeed to a small initial curvature in Planck units, and
then to a large horizon (in Planck length units), allowing a large
homogeneous domain as  initial condition.  
I do not pretend, of course, to provide a final answer to all
problems.   The modest aim  of this
paper is to stress that the problems raised in \cite{1} reduce, in the
end,  to the question of whether the horizon scale,
{\em irrespective of its size},  may be a natural   scale 
for determining the inflationary initial conditions (in particular, the
size of the initial homogeneous region),  and to suggest the possibility
that the answer is not negative ``a priori",  at least when the initial
conditions are imposed well inside the classical regime, like in the case
of the pre-big bang scenario.

Let me  start recalling that the kinematical problems of the
standard scenario can be solved by two classes of accelerated
backgrounds \cite{5}. Consider, for instance, the flatness
problem,  requiring a phase in 
which the ratio $r=k/a^2H^2 \sim \dot a^{-2}$ decreases, so as to
compensate its growth up to the present value $r~ \laq~1$ 
during the subsequent phase of standard evolution. By
parametrizing the scale factor as $a \sim |t|^\b$, the decrease
of $r \sim |t|^{2(1-\b)}$ can be arranged either by 1) $\b > 1$, $t
\ra +\infty$, or 2)  $\b < 1$, $t \ra 0_-$. Both classes of
backgrounds are accelerated, as ${\rm sign}~ {\dot a }= {\rm
sign}~{\ddot a}$.  The first class corresponds
to power-inflation, and includes de Sitter inflation in the limit
$\b \ra \infty$. The second class includes superinflation for $\b
<0$, and accelerated contraction for $0<\b <1$. 

The main kinematic difference between the two classes is the
behaviour of the event horizon, whose proper size is defined by
\beq
d_e (t) = a(t)\int _t^{t_M} dt' a^{-1} (t') .
\label{1}
\eeq
Here $t_M$ is the maximal future extension of the cosmic time
coordinate for the inflationary manifold. Therefore, $t_M=+\infty$ for
the first class, and  $t_M= 0$ for the second class of
backgrounds. In both cases we find that the integral converges,
and that $d_e(t) \sim |H|^{-1}(t)$, so that the horizon size is
constant or growing for class 1), shrinking for class 2), following
the inverse behaviour of the curvature scale.  The phase of
pre-big bang evolution, in particular, is dual  to a phase of standard,
decelerated evolution: its accelerated kinematics is
characterized by a growing curvature scale (i.e. growing $|H|$), and
may be represented as superinflation, in the string frame, or
accelerated contraction, in the Einstein frame \cite{5}. 

In order to recall the criticism of \cite{1} we will now compare the
kinematics of standard de Sitter inflation and pre-big bang
superinflation, for an oversimplified cosmological model in which the
standard radiation era  begins at the Planck scale, and it is
immediately preceeded by a phase of accelerated
(inflationary) evolution.  Also, for the sake of simplicity, we will
identify at the end of inflation 
the present value of the string length $L_s$ 
 with the Planck length $L_p$ (at tree-level, they are
related by $L_p=\langle g\rangle L_s= \langle \exp \phi/2 \rangle
L_s$, with a present dilaton expectation  value $\langle g \rangle
\sim 0.1-0.01$).  

At the beginning of the radiation era the horizon size is thus
controlled by the Planck length $L_p\sim L_s$, while the proper size
of the homogeneous and causally connected region inside our present
Hubble radius, rescaled down at the Planck epoch according to the
standard decelerated evolution of $a(t)$, is unnaturally larger than
the horizon by the factor $ \sim 10^{30} L_p$. During the inflationary
epoch, the ratio \beq
{\rm proper~ size~ horizon~ scale \over
proper ~size~ homogeneous~ region} \sim {H^{-1}(t)\over a(t)} \sim
\eta
 \label{3}
\eeq
must thus decrease at least by the factor $10^{-30}$, so as to
push the homogeneous region outside the horizon, of the
amount required  by the subsequent decelerated evolution. Since
the above ratio evolves linearly in conformal time 
 $\eta \sim \int a^{-1}dt$, the condition
of sufficient inflation can be written as 
\beq
|\eta_f|/|\eta_i| ~\laq ~ 10^{-30}, 
\label{4}
\eeq
where $\eta_i$ and $\eta_f$ mark, respectively,  the beginning and
the end of the inflationary epoch. 

Let us now compare de Sitter inflation, $a \sim (-\eta)^{-1}$, with a
typical dilaton-dominated superinflation, $a \sim
(-t)^{-1/\sqrt 3} \sim (-\eta)^{-1/(\sqrt 3 +1)}$ (the same
discussed in \cite{1}).  

In the standard de Sitter case the horizon and the Planck length are
constant,  $H^{-1} \sim L_s \sim L_p$; as we go back in time, 
according to eq. (\ref{4}), $a(t)$  reduces by the factor
$10^{-30}$ so that, at the beginning of inflation, we find a
homogeneous region just of size $L_p$, like the horizon. 
In the superinflation case, on the contrary,  during the conformal
time interval (\ref{4}), $a(t)$ is only reduced by the factor $a_i/a_f =
10^{-30/(1+\sqrt 3)} \sim 10^{-11}$, so that the size of the
homogeneous region, at the beginning of inflation, is still large in
string units,  $\sim  10^{30\sqrt 3/(1+ \sqrt 3)} L_s \sim 10^{19}
L_s$.   The situation is even worse in Planck units since, at the
beginning of inflation, the string coupling $\exp \phi/2$, and thus
the Planck length $L_p$, are reduced with respect to their final
values by the factor \cite{3} $L_p/L_s=|\eta_f/\eta_i|^{\sqrt 3/2}
\sim 10^{-15 \sqrt 3}$, so that $10^{19} L_s \sim 10^{45}L_p$. This,
by the way, is exactly the initial size of the homogeneous region
evaluated in the Einstein frame in which $L_p$ is constant, and the
above dilaton-driven evolution is represented as a contraction, 
with $a \sim (-\eta)^{1/2}$ (see Fig. 1 for a qualitative illustration
of the differences between de Sitter inflation and pre-big bang
inflation). 

\begin{figure}[t]
\begin{center}
\mbox{\epsfig{file=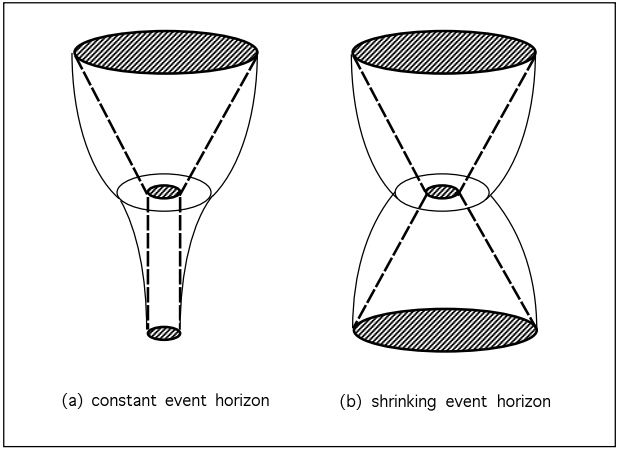,width=82mm}}
\vskip 5mm
\caption{\sl Qualitative evolution of the horizon scale and of
the proper size of a homogeneous region for (a) standard de
Sitter inflation, and (b) pre-big bang superinflation, represented in
the Einstein frame as a contraction.  The time direction coincides with
the vertical axis. The three horizontal spatial sections corresponds,
from top to bottom, to the present time, to the end and to the
beginning of inflation. The shaded area represents the horizon,  and
the dashed lines its time evolution. The full curves represent the time
evolution of the border of the homogeneous region, controlled by 
the scale factor.}  
\end{center}
\end{figure}

According to \cite{1}, case (a) of Fig. 1
provides an acceptable example of inflationary scenario, as the
initial  homogeneity scale is contained within a single domain of
Planckian size.  Case (b), on the contrary, is not satisfactory
because of the initial  homogeneity on scales much greater than
Planckian, $ 10^{19} L_s \sim 10^{45}L_p$.  Quoting Ref. \cite{1}, this
situation ``is not much better than  the situation in the
non-inflationary big bang cosmology, where it was  necessary to
assume that the initial size of the homogeneous part of  our
Universe was greater than $10^{30}L_p$".

I would like to stress, however, that in 
case (b) the initial homogeneous region is
large in Planck  units, {\em but not larger than the horizon itself}. 
Indeed, during superinflation, the horizon scale 
shrinks linearly in cosmic time. As we go backwards in time, for
the particular example that we are considering, the horizon
increases by the factor $H^{-1}_i/H^{-1}_f=|t_i|/|t_f|=
(\eta_i/\eta_f)^{\sqrt 3/(1+ \sqrt3)}$, so that, at the beginning
of inflation, $H^{-1} \sim 10^{30  \sqrt 3/(1+\sqrt3)}L_s\sim
10^{19}L_s\sim 10^{45}L_p$, i.e.  the horizon size is just the same as
that of the homogeneous region (as illustrated in Fig. 1). In this
sense, both initial conditions, in cases (a) and  (b), seem to be
equally natural. The difference is that in case (b) the
initial horizon is large in Planck units, while in case (a) it is of
order one. This is an obvious consequence of the different
curvature scales at the beginning of inflation. 

The question about the naturalness of the initial
conditions seems thus to concern the unit of length used, in
particular,  to measure the size of the initial homogeneous domain,
and, more generally,  to characterize the initial geometric
configuration at the onset of inflation: which basic length scale has
to be used, the Planck (or string) length, or the radius of the causal
horizon? 

This, I believe, is the question to be answered. 
Providing a definite answer may deserve a careful analysis,
which is outside the scope of this brief paper. Let me note that,
according to \cite{1}, it is the Planck (or string) 
scale that should provide the natural units for the size of the initial
homogenous patches and for the initial curvature and coupling scale. 
This is certainly reasonable when initial conditions are imposed on a
cosmological state approaching the high-curvature, quantum gravity
regime. In the pre-big bang scenario, however, initial conditions are
to be imposed when the Universe is deeply inside the low-curvature,
weak coupling, classical regime. In that regime the Universe does not
know about the Planck length, and the  causal horizon $H^{-1}$ could
represent a natural candidate for controlling the set of initial
conditions. For what concerns homogeneity, however, I am not
suggesting that the horizon (which is the maximal homogeneity scale) 
should be always {\em assumed} as the natural scale of homogeneity. I
am suggesting that this possibility should be discussed on the ground of
some quantitative and objective criterium, as attempted for instance in
\cite{6}, and not discarded a priori, as in \cite{1} (see also \cite{7a} for
a discussion of ``generic" initial conditions in a string cosmology
context). 

One might think that, accepting the horizon size as a  natural
homogeneity scale, there is no need of inflation to explain our
present homogeneous Universe \cite{1}.  This is not the case,
however, because if we go back in time  without inflation our
Universe should start in the past from a homogeneous region
unnaturally larger than the horizon (see Fig. 1). Only with  inflation
the homogeneous region, going back in time, re-enters inside the
horizon. So, only if there is inflation, an initial homogeneity scale of
the order of the horizon scale is enough to reproduce our present
Universe. 

Also, one might think,  as noted in \cite{1}, that the classical
homogeneity of  the horizon might be destroyed by quantum
fluctuations amplified  during the contraction preceeding  the onset
of the inflationary era,   in such a way as to prevent the
formation of a large homogeneous domain. This problem has been
recently discussed in \cite{7} for the case of a homogeneous string
cosmology background with negative spatial curvature: it has been
shown that quantum fluctuations die off much faster than classical
inhomogeneities as they approach the initial perturbative vacuum,
and remain negligible throughout the perturbative pre-big bang
phase.  For classical perturbations, however, the situation is different,
and no general result is presently available. The initial amplitude of the
classical inhomogeneities is not normalized to a vacuum fluctuation
spectrum, the results of \cite{7} cannot be applied, and inflation can
occurr successfully or not depending on the initial distribution of the
classical amplitudes.

Finally, one might argue that a large initial horizon, assuming a
saturation of the bound imposed by the holographic principle in a
cosmological context \cite{8}, implies a large initial entropy, $S=$
(horizon area in Planck units), and thus a small probability for the
initial configuration. Indeed, if $S$ is large, the probability that such 
a configuration be obtained through a process of quantum tunnelling 
(proportional to $\exp [-S]$) is exponentially suppressed, as
emphasized in \cite{1}. However, in the pre-big bang scenario,
quantum effects such as tunnelling or reflection of the Wheeler-De
Witt wave function are expected to be important
towards {\sl the end} of inflation \cite{9}, and  {\sl not the
beginning}, as  they may be effective 
 {\sl to exit} \cite{10}, eventually, from the inflationary regime, 
 {\sl not to enter} it and
 to explain the origin of the initial state. A
large entropy of  the initial state, in the weakly coupled, highly
classical regime, can  only correspond to a large probability of such
configuration,  (proportional to $\exp [S]$), as expected for classical
and macroscopic  configurations. 

In conclusion, let me come back  on the large
dimensionless parameters characterizing the initial state of pre-big
bang inflation \cite{1}. The physical meaning of those parameters,
i.e. the fact that the initial string coupling and curvature are very
small in string (or Planck) units, is to be understood as a
consequence of the perturbative initial conditions, 
suggested by the underlying duality symmetries. On the other hand,
whenever inflation starts at curvature scales smaller than
Planckian, the initial state is necessarily characterized by a large
dimensionless ratio -- the inverse of the curvature in Planck units.
If one believes that such large numbers should be avoided,  then
should be prepared to accept the fact that natural initial conditions
are only possible in the context of models in which inflation starts at
the Planck scale: for instance chaotic inflation, as pointed out in
\cite{1}. 

This is a rather strong conclusion, that rules out, as a satisfactory
explanation of our present cosmological state, not only the pre-big
bang scenario, but any model in which inflation starts at scales
smaller than Planckian (unless we have a scenario with different
stages of inflation responsible for solving different problems). 
Even for a single stage of inflation very close to the Planck scale, 
however, we are not free of problems, as we are led, eventually, to 
the following question: can we trust the naturalness of inflation models
like chaotic inflation, in which classical general 
relativity is applied to 
set up initial conditions at Planckian curvature scales, i.e. deeply
inside the non-perturbative, quantum gravity regime? 

The Planckian regime is certainly problematic to deal with, both in the
string and in the standard inflationary scenario: in string cosmology, in
particular, it prevents a simple solution of the ``graceful exit" problem 
\cite{11}.  The pre-big bang scenario, however, tries to look back
in time beyond the Planck scale by using the powerful tools of
superstring theory, in particular its duality symmetries.  According to
duality, the pre-Planckian Universe approaches initially the state of a
low-energy system, and initial conditions are to be set up in a regime
well described by the lowest order effective action, in which all
quantum and higher-order corrections are small,  and under control. 
It is true, however, that the presence of the Planckian regime can 
indirectly affect the initial conditions also in a string cosmology 
context, as it imposes a finite duration of the low-energy 
dilaton-driven phase: the initial homogeneity scale, as a consequence, 
has to be large enough to emerge with the required size at the 
Planck epoch, and to avoid the need for a further period of 
high-curvature, Planckian inflation \cite{13a}. 

It should be stressed, finally,  that the  main
difference from the standard scenario, in which any tracks of the
pre-Planckian cosmological state is washed out by inflation, is probably
the fact that the pre-Planckian history may become visible, in the sense
that its phenomenological consequences can be tested (at least in
principle) even today \cite{12}. So, while in the context of standard
inflation the naturalness criterium can be safely applied to select an
initial state at the Planck scale, it seems difficult (in my opinion) to
apply the same criterium in a string cosmology context, and to discard a
model of pre-Planckian evolution only on the grounds of the large
parameters characterizing the initial conditions. Such initial conditions
have consequences accessible to observational tests, and the analysis
of the ``a posteriori" probabilities with the Bayesan approach of \cite{6}
suggests that a state with a large initial horizon may become ``a
posteriori" natural, because of the duality symmetries intrinsic to the
pre-big bang scenario. 

However, much further work is certainly needed before a final
conclusion is reached. Irrespective of the final results, such work will
certainly improve our present understanding of string theory and of
the physics of the early Universe.

\acknowledgments
I wish to thank Raphael Bousso, Nemanja Kaloper, Andrei Linde, 
and Gabriele Veneziano for stimulating discussions and helpful 
comments (not necessarily in agreement with the personal point of
view presented in this paper).

\end{document}